\documentclass[conference]{IEEEtran}
\IEEEoverridecommandlockouts
\usepackage{cite}
\usepackage{amsmath,amssymb,amsfonts}
\usepackage{algorithmic}
\usepackage{graphicx}
\usepackage{textcomp}
\usepackage{xcolor}
\usepackage{color}
\usepackage{booktabs}
\usepackage{xspace}
\usepackage{multirow}
\usepackage[framemethod=TikZ]{mdframed}
\usepackage{colortbl}
\usepackage{ragged2e}
\usepackage{courier}
\usepackage{listings}
\usepackage{varwidth}
\usepackage{newfloat}

\def\BibTeX{{\rm B\kern-.05em{\sc i\kern-.025em b}\kern-.08em
    T\kern-.1667em\lower.7ex\hbox{E}\kern-.125emX}}

\usepackage[normalem]{ulem}

\newcommand{\myparagraph}[1]{\vspace{0.1cm}\noindent \xspace\textbf{\textit{#1.}}}

\newcommand{\ie}{i.e.,\xspace}
\newcommand{\eg}{e.g.,\xspace}

\newcommand{\etal}{et al.\xspace}

\newcommand{\seclabel}[1]{\label{sec:#1}}

\newcommand{\figref}[1]{Figure~\ref{fig:#1}}
\newcommand{\secref}[1]{Section~\ref{sec:#1}}

\newboolean{showcomments}
\setboolean{showcomments}{true}
\ifthenelse{\boolean{showcomments}}
	{\newcommand{\nb}[3]{
		{\colorbox{#2}{\bfseries\sffamily\scriptsize\textcolor{white}{#1}}}
		{\textcolor{#2}{\sf\small$\blacktriangleright$\textit{#3}$\blacktriangleleft$}}}
	 }
	{\newcommand{\nb}[3]{}
	 }

\usepackage{xcolor}
\usepackage{textcomp}
\usepackage[T1]{fontenc}
\usepackage{inconsolata}

\newcommand{\ct}{\lstinline[backgroundcolor=\color{white},basicstyle=\footnotesize\ttfamily]}

\definecolor{pblue}{rgb}{0.13,0.13,1}
\definecolor{pgreen}{rgb}{0,0.5,0}
\definecolor{pred}{rgb}{0.9,0,0}
\definecolor{pgrey}{rgb}{0.46,0.45,0.48}
\definecolor{lightgreen}{rgb}{0.56, 0.93, 0.56}
\definecolor{lightgray}{rgb}{0.83, 0.83, 0.83}

\begin{document}

\title{A manual categorization of new quality issues on automatically-generated tests}


\author{\IEEEauthorblockN{Geraldine Galindo-Gutierrez}
\IEEEauthorblockA{\textit{Bolivian Catholic University, Bolivia} \\
}
\and
\IEEEauthorblockN{Maxilimiliano Narea}
\IEEEauthorblockA{\textit{Pontificia Universidad Católica de Chile, Chile} \\
}
\and
\IEEEauthorblockN{Alison Fernandez Blanco}
\IEEEauthorblockA{\textit{Pontificia Universidad Católica de Chile, Chile} \\
}
\and
\IEEEauthorblockN{Nicolas Anquetil}
\IEEEauthorblockA{\textit{Université de Lille, France} \\
0000-0003-1486-8399}
\and
\IEEEauthorblockN{Juan Pablo Sandoval Alcocer}
\IEEEauthorblockA{\textit{Pontificia Universidad Católica de Chile, Chile} \\
0000-0002-8335-4351}
}

\definecolor{dkgreen}{rgb}{0,0.6,0}
\definecolor{gray}{rgb}{0.5,0.5,0.5}
\definecolor{mauve}{rgb}{0.58,0,0.82}
\definecolor{light-gray}{gray}{0.97}  
\definecolor{weborange}{rgb}{255,165,0}
\definecolor{darkviolet}{rgb}{0.5,0,0.4}
\definecolor{darkgreen}{rgb}{0,0.4,0.2} 
\definecolor{darkblue}{rgb}{0.1,0.1,0.9}
\definecolor{darkgrey}{rgb}{0.5,0.5,0.5}
\definecolor{lightblue}{rgb}{0.4,0.4,1}

\definecolor{stringColor}{rgb}{0.16,0.00,1.00}
\definecolor{fieldColor}{rgb}{0.16,0.00,1.00}
\definecolor{annotationColor}{rgb}{0.39,0.39,0.39}
\definecolor{keywordColor}{rgb}{0.50,0.00,0.33}
\definecolor{commentColor}{rgb}{0.25,0.50,0.37}
\definecolor{javadocColor}{rgb}{0.25,0.37,0.75}
\definecolor{jTagColor}{rgb}{0.50,0.62,0.75}
\definecolor{eTagColor}{rgb}{0.50,0.62,0.75}
\definecolor{lineNumberColor}{rgb}{0.47,0.47,0.47}
\definecolor{shadecolor}{rgb}{0.9,0.9,0.9}

\renewcommand{\lstlistingname}{Fig.}
\newcommand\coderef{Figure~\ref}
\lstset{ %
  language=Java,                
  basicstyle={\scriptsize\ttfamily},     
  numbers=left,               
  linewidth=0.98\linewidth,
  xleftmargin=2.5em,
  backgroundcolor=\color{light-gray},  
  showspaces=false,             
  showstringspaces=false,         
  showtabs=false,                 
  frame=single,                   
  rulecolor=\color{black},       
  tabsize=2,                   
  captionpos=b,        
  breaklines=true,     
  abovecaptionskip=2pt,
  breakatwhitespace=false,       
  title=\lstname,                                                  
  commentstyle=\color{commentColor},    
  stringstyle=\color{stringColor},  
  escapeinside={\%*}{*)},
  keywordstyle = {\color{keywordColor}}
}
  

\newcounter{rmd}
\newenvironment{rmd}[1][]{
    \refstepcounter{rmd}
    \mdfsetup{innertopmargin=0.7Em,
              linecolor=gray!20,
              linewidth=2pt,
              topline=true, 
              backgroundcolor=light-gray,
              nobreak=true}
    \begin{mdframed}[] {\bf Recommendation~\thermd.} \relax
    }{\end{mdframed}}

\maketitle

\begin{abstract}
Diverse studies have analyzed the quality of  automatically generated test cases by using  test smells as the main quality attribute.
But recent work reported that generated tests may suffer a number of quality issues not necessarily considered in previous studies.
Little is known about these issues and their frequency within generated tests.
In this paper, we report on a manual analysis of an external dataset consisting of 2,340 automatically generated tests.
This analysis aimed at detecting new quality issues, not covered by past recognized test smells.
We use thematic analysis to group and categorize the new quality issues found.
As a result, we propose a taxonomy of 13 new quality issues grouped in four categories.
We also report on the frequency of these new quality issues within the dataset and present eight recommendations that test generators may consider to improve the quality and usefulness of the automatically generated tests.
\end{abstract}


\section{Introduction}

Unit tests are a critical asset in software development, as they help identify and resolve issues early in the development cycle.
By verifying the behavior of individual program units in isolation, unit tests help ensure that each unit of code performs as intended, saving developers significant time and resources.
Unit tests can be created either manually or automatically. 
Manually created tests can be time-consuming and may only cover a limited set of test scenarios that developers or testers consider. 
For this reason, diverse test  strategies have been proposed to automatically generate tests and reduce manual effort. 
The latter can complement the manually created test battery, increasing coverage and fault detection.

To ensure the effectiveness of unit tests, they must meet several quality attributes, such as coverage, readability, maintainability, isolation, and speed. 
Previous studies have used test smells to detect generated tests that contain code pieces that may compromise their quality~\cite{peruma2020tsdetect,panichella2022,grano2019scented}. 
For example, it has been found that automatically generated tests often exhibit typical test smells, such as Eager Testing or Assertion Roulette. 
However, Panichella \etal~\cite{panichella2022} recently identified additional issues not fully captured by test smells. 
For instance, tests that lack assertions and tests that involve a substantial amount of setup but still result in exceptions. 
Consider an example we found in our manual inspection, as shown in \figref{aa-mistmach-example2}. 
In this case, the test assertions (lines 5--6) are only related to the object initialization (lines 2--3), and the method call \ct{setStartDate} (line 4) is unrelated to the rest of the test. 
These issues are partly associated with the unique nature of automatically generated tests and reflect a disconnect between the optimization metric of "coverage" and the real-world validity of test cases. 
While previous studies reported three of these issues, it remains unclear whether there are additional similar issues and how widespread they are in automatically generated tests.

\begin{figure}
 \begin{lstlisting}[language = Java, escapechar=^ , label={fig:aa-mistmach-example2}, caption = Example of test with inconsistent assertions.]
public void test28() throws Throwable {
  XmlCalendar xmlCalendar0 = new XmlCalendar();
  CronText cronText0 = new CronText(xmlCalendar0);
  cronText0.setStartDate(xmlCalendar0);
  assertEquals(1, cronText0.getFrequency());
  assertEquals(0, cronText0.getInterval());
}
\end{lstlisting}
\end{figure}

In this paper, we manually review 2,340 automatically generated tests provided in the curated dataset of Panichella and colleagues~\cite{panichella2020evosuite} to detect quality issues that do not fully match previously studied test smells. 
Similar to previous work, we use the term ``issue'' to refer to portions of code that reflect symptoms or concerns related to a broader quality problem, where ``issue'' is the closed vocabulary to refer to them~\cite{panichella2022}. 
We refer to our proposition as \emph{quality issues}, whereas we use the phrase \emph{test smells} for smells that have been published previously.

We investigated the quality issues  through an empirical study. 
First  (\secref{Smells}), two of the authors manually reviewed the curated dataset to locate (already known) test smells and ensure that the (new) quality issues were not already captured by any test smells. 
Then (\secref{Thematic}), they detected and categorized ``issues'' by conducting a thematic analysis and holding four meetings to refine the codes and themes generated. 
The remaining authors held three meetings to discuss disagreements to check the consistency of the thematic analysis. 
Finally, we manually analyzed the ``issues'' and counted the frequency of issues for projects, classes, and methods within the curated dataset. 
As a result, we developed a taxonomy with 13 categories and eight recommendations that test generator maintainers and researchers can consider to improve or use test generator tools. 
The paper's main contribution is the design and execution of an empirical study that analyzes quality issues not fully covered by current test smells.

\section{Related Work}\label{sec:RelatedWork}
This section summarizes the literature related to (i) test smells, (ii) automatic test generation, (iii) test smells in generated tests. It also points out the limitations of the prior work.

\myparagraph{Test Smells} Test smells extend the definition of code smells~\cite{fowler1999,cairo2018impact,palomba2018diffuseness} to tests (\ie, structures or patterns that indicate lousy test code~\cite{deursen2001}). Deursen \etal~\cite{deursen2001} presented the first set of test smells in 2001, and posterior studies involving other authors~\cite{peruma2019,meszaros2003} expanded the collection. By 2018, Garousi \etal~\cite{garousi2018} found eight formally published sources presenting new test smells~\cite{chen2012,deursen2001,dudney2003,freeman2009,jenitta2002,lundin2015,meszaros2007,page2008}. Aljeedaani \etal ~\cite{baker2006,breugelmans2008testq,delplanque2019,greiler2013automated,gyori2015reliable,huo2014improving,koochakzadeh2010tecrevis,peruma2020tsdetect} studied test smells detection tools and found that nine presented new test smell definitions. Many empirical studies~\cite{bavota2012,tufano2016} have investigated the prevalence of test smells and their effects and found them to be detrimental to program comprehension and maintenance. As a result, some of these test scents are used for analyzing automatically generated tests (see \secref{SmellsGeneratedTests}) and for improving the quality of generated tests.

\subsection{Automatically-Generated Tests}
The goal of automatic test generation is to save effort and time by generating test suites with minimal human intervention~\cite{bacchelli2008,fraser2015does,kracht2014,Rojas2015}. Among various generation tools~\cite{csallner2004,Tonella2004,Pacheco2007}, our focus in this study is on the EvoSuite Java tool~\cite{Fraser2011,panichella2020evosuite}. EvoSuite is designed to achieve the maximum possible coverage on its generated test suites~\cite{Fraser2012} using generation algorithms that consider the test generation as a multi-objective problem~\cite{panichella2017automated,panichella-2018-dynamosa}. With its latest algorithm, DynaMOSA, EvoSuite produces shorter tests with higher coverage~\cite{campos2018,panichella-2018-dynamosa,panichella2022}. In order to improve EvoSuite, a number of studies have focused on: (i) readability~\cite{Grano2018,lin-quality-indentifiers,Daka2015,panichella2016impact,roy2020,daka2017generating} and (ii) scenario generation~\cite{serra2019,shamshiri2018}.

\subsection{Test Smells in Automatically-Generated Tests}\seclabel{SmellsGeneratedTests}
Previous studies have explored test smells in generated tests using detection tools(\eg tsDetect~\cite{peruma2020tsdetect}, JNose~\cite{virginio2020jnose}) and manual revision~\cite{panichella2020evosuite}. Palomba \etal ~\cite{palomba2016diffusion} used the Test Smell Detector~\cite{bavota2015test} to investigate the frequency of the Deursen \etal test smells~\cite{deursen2001} on the SF110 dataset~\cite{Fraser2014}. The results showed that 83\% of the classes were smelly. Grano \etal conducted a similar study with a smaller but significant sample of SF110~\cite{grano2019scented}. The results showed a high proportion of tests generated with Assertion Roulette and Eager Test smells. Panichella \etal~\cite{panichella2022} presented a new review of the topic analyzing the Deursen smells~\cite{deursen2001}, using the same sample as Grano \etal~\cite{grano2019scented}, but generating tests with DynaMOSA algorithm~\cite{panichella2017automated} and extended parameters in EvoSuite. As a result, Panichella \etal found that such test smells were less extensive than thought when tests were generated with better configurations.


\myparagraph{Limitations in the Prior Work} 
Most previous studies examine the quality of generated tests by focusing on test smells using detection tools and manual inspection. However, these studies voluntarily ignore other problems present in generated tests. Recently, Panichella and colleagues~\cite{panichella2022} reported ``issues'' that are not necessarily fully captured by current test smells. For example, they found tests with too many assertions, and tests that required significant setup that ended up throwing exceptions indicating failed setup. There is no thorough investigation of these problems or atypical code patterns within the generated tests. Therefore, little is known about their frequency, impact, or the existence of other problems.

Compared to most of the literature, our study is focused on locating, exploring, and quantifying ``issues'' that (i) affect the quality of automatically generated tests \cite{brandt2022, panichella2022} and (ii) are not necessarily captured by the current test smells~\cite{panichella2022}.

\section{Methodology} \seclabel{methodology}
To identify and quantify quality issues~\cite{brandt2022} that do not fully match with existing test smells~\cite{panichella2022} in generated tests, we focus on the following research questions:

\begin{itemize}
    \item \textbf{$RQ_1$} \emph{-- What issues or atypical patterns are present in automatically-generated tests? }
    \item \textbf{$RQ_2$} \emph{-- How widespread are these issues in automatically-generated tests?}
\end{itemize}

 \textbf{RQ1} centers on discovering and categorizing quality issues in automatically-generated tests. These issues (i) affect the quality of generated tests \cite{brandt2022} and (ii) are not necessarily represented by current test smells~\cite{panichella2022}. To answer \textbf{RQ1}, we opted for an approach of two steps: (i) {collecting existing test smells in automatically-generated tests }(described in \secref{Smells}), and (ii) manually locating issues and generating a classification scheme based on the detected issues (described in \secref{Thematic}). We also provide recommendations to improve the quality and usefulness of automatically generated tests based on the detected issues. Finally, \textbf{RQ2} focuses on quantifying the identified issues. To answer \textbf{RQ2}, we manually analyze the issues and collect the frequency of issues for projects, classes, and methods.

The following subsections detail the steps in our methodology.

\subsection{Dataset Selection}\label{sec:Dataset}
To detect and quantify the  quality issues in automatically-generated tests, we chose the curated dataset of Panichella and colleagues~\cite{panichella2022}. We opted for this dataset because (i) it contains non-trivial classes sample suitable for studies centered on understanding and characterizing automatically-generated tests~\cite{Shamshiri2015, grano2019scented}, and (ii) it involves tests generated with better configurations using the DynaMOSA algorithm~\cite{panichella2017automated}, and enhanced hyperparameters. 

\subsection{Test Smells {Review}}\label{sec:Smells}
To identify quality issues that may not be captured by the existing test smells definitions, we conducted a thorough review of the definitions of test smells considered in previous studies focused primarily on automatically generated tests\cite{panichella2022, peruma2019, Grano2018, deursen2001}. We used this information as a starting point for our manual inspection process, which allowed us to analyze the code and detect any issues not covered by the current test smells.



\subsection{Quality Issues Identification and Classification}\label{sec:Thematic}
We performed a thematic analysis \cite{Braun2017} to detect and classify quality issues. {To locate quality issues, the reviewer should detect an issue that affects a test case's quality and examine that this issue is not captured by a test smell.} Consequently, the current test smells could not be reported as new quality issues.


In order to create a classification scheme and locate quality issues in automatically-generated tests, two authors conducted a thematic analysis following a number of specific steps detailed in the next paragraphs.

\myparagraph{Familiarization} The two authors independently read and reread each test case to have an overview of the structure and the information of the test cases in the dataset.

\myparagraph{Generating codes} {The two authors independently determine if any quality issue in each test case was not captured by existing test smells in \secref{Smells}. If an author considers a quality issue in the source code, reviews the test smells' definition, and notes that the located quality issue does not match any current test smell.} Then, the author makes an annotation where he/she assigns codes that reflect the relevant features of the detected quality issue and mentions the variable or method involved with the issue. For example, one author assigned the code \emph{duplicated\_setup} to indicate that in a test case, two or more code lines are equally exact from the setup in common. Additionally, the two authors conducted two continuous reviews to refine codes and determine if they were assigned correctly. The latter requires comparing two test cases assigned to the same code to inspect if they reflect the same issue. 

\myparagraph{Constructing initial themes} To generate coherent groups, the two authors independently compiled the assigned codes along their associated test cases. These groups allow identifying initial themes (broader patterns) that help address the research questions. The codes that did not belong to a specific theme were grouped as miscellaneous and analyzed in the next step.

\myparagraph{Reviewing themes} The two authors held two meetings to check the initial themes created previously against the associated test case (\eg lines of code with the issue and annotations). Then, they refined the initial themes to create a final set of themes.

\myparagraph{Defining and naming themes} Each theme of the final set was defined with a description and an informative name based on the identified patterns. For example, the theme \emph{Redundant code} contains quality issues involved with duplicated, redundant, and/or unnecessary test code that can be extracted to avoid redundancy or erased for simplicity. Note that this theme is related to the test smell duplicated code. However, each quality issue presents features that do not precisely match the exact definition of the test smell duplicated code. Consequently, our quality issues illustrate this difference in the interpretation rules and corresponding examples.

Finally, to ensure the consistency of the process, the remaining authors checked the consistency of codes and themes against the associated data. They examined if the themes were created to respond to the research questions. Three meetings were held to discuss the disagreements or potential issues of the generated codes and themes. As a consequence, we minimized potential inconsistencies in the coding process.

Note that the identified quality issues can be represented as patterns in test cases. Therefore, they can be used to improve the quality of test generators.


\section{Results}
We manually review 2,340 automatically generated tests to detect and quantify quality issues. A total of two meetings were held to identify test smells, and seven meetings were carried out to classify and locate quality issues. The results of our empirical study are detailed in the following subsections.

\subsection{Quality Issues (RQ1)}
After manually inspecting and categorizing the data using thematic analysis, we identified a catalog of 13 quality issues that we grouped into four categories. Table \ref{tab:issues-discovered} summarizes these categories, their rationale, as well as the rules (symptoms) we used to manually detect and collect the frequency. This section details each of the categories we found, providing examples and a number of recommendations to improve the test generator tools.


\begin{table*}[ht!]
\centering
\setlength{\tabcolsep}{8pt}
\renewcommand{\arraystretch}{1.5}
\caption{Observed quality issues in automatically generated tests}
\label{tab:issues-discovered}
\begin{tabular}{p{0.15\textwidth}|p{0.35\textwidth}|p{0.40\textwidth}}
\hline
\multicolumn{3}{c}{\cellcolor[HTML]{EFEFEF}\textbf{Act-Assert Mismatch}} \\ \hline
Not asserted side effects (NASE) 
& \textbf{Symptom.} At least one void method call causes a side effect, but no assertion verifies the side effect or references the method call.
& \textbf{Rationale.} Although a void method call may increase coverage if no assertion is associated, the test reduces its ability to detect future change behavior rather than exceptions.
\\ \hline
Not asserted return values (NARV)
& \textbf{Symptom.} At least one method call returns a value that is not checked by any assertion or used later in the test.
& \textbf{Rationale.} If a test excludes method call return values from its assertions, it misses the opportunity to verify that the behavior of the method under test is correct.
%
\\  \hline
Assertions with not related parent class method (ARPM)
& \textbf{Symptom.} At least one assertion calls an inherited method that checks a value unrelated to the method under test.
& \textbf{Rationale.} Asserting methods that are not tested is not very useful. Also, the peculiarity of asserting parent class methods may fit as a special case of indirect testing.
\\ \hline

\multicolumn{3}{c}{\cellcolor[HTML]{EFEFEF}\textbf{Redundant Code}} \\ \hline
Asserting object initialization multiple times (OIMT)
& \textbf{Symptom.} Two or more tests in which the assertions check for values that are set in the constructor, or check for default values.
& \textbf{Rationale.} A limited number of tests should be dedicated to checking the correct initialization of an object. However, this becomes redundant when repeated throughout the suite.
\\ \hline
Duplicated Setup (DS)
& \textbf{Symptom.} Two or more tests sharing at least two equal lines of setup code.
& \textbf{Rationale.} Test maintenance and analysis becomes difficult due to duplicate code.
\\ \hline
Testing the same exception scenario (TSES)
& \textbf{Symptom.} Two or more tests with the same exceptions and console output calling different methods with the same setup.
& \textbf{Rationale.} If there are two tests that evaluate the same exception scenario, one of them can be considered to be semantically redundant.
\\ \hline
Testing the same void method (TSVM)
& \textbf{Symptom.} Two or more tests that call the same void method and that report instances of side effects that are not asserted.
& \textbf{Rationale.}  It is important to have two or more tests of the same method to cover different branches. But their usefulness for finding bugs is limited if the assertions are unrelated to the behavioral differences in the method.
\\ 
\hline

Redundant not null Assertion (NNA)
& \textbf{Symptom.} At least one assertNotNull is redundant. It is redundant (i) after initializing an object, (ii) if some other assert checks the value content, and (iii) if the method being tested does not return null.
& \textbf{Rationale.} In order to reduce the size of the test and the amount of time that developers spend analyzing it, it may be useful to remove redundant statements.
\\ \hline
\multicolumn{3}{c}{\cellcolor[HTML]{EFEFEF}\textbf{Failed Setup}} \\ \hline
Exceptions due to null arguments (EDNA)
& \textbf{Symptom.} Tests handling NullPointerException caused by null arguments passed directly to constructors or methods.
& \textbf{Rationale.} In some scenarios, testing for exceptions caused by null arguments can be valuable. However, if the class is designed to assume that no null arguments are sent, testing will not cover new branches, and the result is predictable.
\\ \hline
Exceptions due to external dependencies (EDED)
& \textbf{Symptom.} Tests that handle \texttt{HeadlessException}, \texttt{SQLException} or \texttt{NotYetConnectedException}.
& \textbf{Rationale.} Typically, these exceptions occur when a test uses external resources. Thus, these tests are no longer self-contained.
\\ \hline
Exceptions due to incomplete setup (EDIS)
& \textbf{Symptom.} Tests that construct an object, call a method and handle an exception thrown by an uninitialized value.
& \textbf{Rationale.} In these cases, much of the generated suite contains exceptions caused by incomplete test setup steps. Rather than interesting test scenarios, these tests reflect the difficulty of the generator tool.
\\ \hline

\multicolumn{3}{c}{\cellcolor[HTML]{EFEFEF}\textbf{Testing only Field Accessors and Constants}} \\ \hline

Testing only field accessors  (TOFA)
& \textbf{Symptom.} Tests containing only object initialization and field accessor (getter/setter) assertions.
& \textbf{Rationale.} Field accessors are straightforward and predictable operations; we expect an obvious result upon their assertion.
\\ \hline
Asserting constants (AC)
& \textbf{Symptom.} At least one assertion verifies the value of a variable declared as a static final on its class.
& \textbf{Rationale.} Asserting constants is uncommon because the assert won't fail unless the value is manually changed.
\\ 
\bottomrule
\end{tabular}
\end{table*}

\subsubsection{Act-Assert Mismatch}
The \emph{Arrange-Act-Assert} (AAA) pattern is a standard structure for unit tests. In this pattern, the Arrange step involves setting up the necessary objects and data to perform the test, the Act step calls the stimulus (i.e., the methods or actions being tested), and the Assert step verifies the output of the stimulus. During our manual analysis, we found that some tests included method calls unrelated to the assertions made.

\myparagraph{Not Asserted Side Effect} This category group tests that contain calls to void methods that cause a side-effect (\ie  they change the object's state), but the assertions within the test do not verify these effects. For example, consider the generated test for the class \ct{SubstringLabeler} shown in \coderef{fig:not_asserted_side_effect_example}. This test only calls the void method \ct{addDataSourceListener}, which adds a \ct{TextViewer} instance to an internal collection of listeners within the \ct{SubstringLabeler} object. However, the assertions within the test only verify attributes such as \ct{matchAttributeName} and \ct{customName}, which are unrelated to the void method call. Although the void method call increases test coverage, the test itself does not help to verify the correct behavior of the \ct{addDataSourceListener} method.

\begin{figure}
\begin{lstlisting}[language = Java, escapechar=^, caption = Not asserted side effect. , label={fig:not_asserted_side_effect_example}, ]
public void test79() throws Throwable {
 SubstringLabeler ^\color{darkgrey}{substringLabeler0}^ = new SubstringLabeler();
 TextViewer ^\color{darkgrey}{textViewer0}^ = new TextViewer();
 ^\color{darkgrey}{substringLabeler0}^.^\textbf{addDataSourceListener}^(^\color{darkgrey}{textViewer0}^);
 ^\textit{assertEquals}^("Match",
            ^\color{darkgrey}{substringLabeler0}^.getMatchAttributeName());
 ^\textit{assertEquals}^("SubstringLabeler", 
            ^\color{darkgrey}{substringLabeler0}^.getCustomName());
}
\end{lstlisting}
\end{figure}

We found 328 (14\%) tests that contain this code pattern. One question that arises for these tests is: \emph{Is it possible to verify these side effects?} After manually reviewing the classes under test, we realized that 31\% of the side-effects can be verified by calling another method within the class, but 69\% of the classes do not provide any public method (\ie accessors) to help verify these side-effects (e.g., printing values in a terminal). We found only one test that contains both cases.


\myparagraph{Not Asserted Return Values} This category group tests that contain a method call that returns a value not verified by the test's assertions. For instance, consider the generated test for the class \ct{ExternalXid} (\coderef{fig:not_asserted_return_values}), where none of its assertions verify the returned value of the \ct{equals} method call. We inspected the \ct{equals} method and confirmed that: (i) it returns a \ct{boolean}, (ii) it does not cause any side effects, and (iii) it has no relation with the \ct{formatId} (the only assertion in the test). Consequently, we concluded that although this increases the coverage, it does not help verify this method's behavior, as it is the only method being tested. We found 242 (10.3\%) tests in this category.

\begin{figure}
\begin{lstlisting}[language = Java, escapechar=^, frame = single, caption = Example of not asserted return values., label={fig:not_asserted_return_values}]
public void test22() throws Throwable {
  byte[] ^\color{darkgrey}{byteArray0}^ = new byte[6];
  ExternalXid ^\color{darkgrey}{externalXid0}^ = new ExternalXid((-1), ...);
  Object ^\color{darkgrey}{object0}^ = new Object();
  ^\color{darkgrey}{externalXid0}^.^\textbf{equals}^(object0);
  assertEquals((-1), ^\color{darkgrey}{externalXid0}^.getFormatId());
}
\end{lstlisting}
\end{figure}

\myparagraph{Assertions with not related parent class method} We found that 121 (5.1\%) tests contain assertions that use methods from the parent class of the class under test. However, these methods are not related to the methods being tested. For instance, consider the following generated test that asserts the method \ct{isFocusTraversalPolicySet} defined in the \ct{Container}class, although the class under test is \ct{Field} (\coderef{fig:assertions_with_not_related_parent_class_method}). 
The \ct{Container} class is extended by the \ct{JComponent} class, which is then extended by the \ct{JPanel} class, and finally by the \ct{Field} class. 
It is also important to note that we verified that the method \ct{isFocusTraversalPolicySet} is not overridden by the \ct{Field} class. In this example, the \ct{paintComponent}method defined in \ct{Field} does not modify the parent class attribute \ct{focusTraversalPolicy}. 
Therefore, the assertion that calls the method \ct{isFocusTraversalPolicySet} is not related to the \ct{paintComponent} method, which is the only operation called after object initialization.

\begin{figure}
\begin{lstlisting}[language = Java, escapechar=^ ,label ={fig:assertions_with_not_related_parent_class_method}, caption = Example of assertions with not related parent class method.]
public void test12() throws Throwable {
  ...
  Field ^\color{darkgrey}{field0}^ = new Field(^\color{darkgrey}{handballModel0}^, ^\color{darkgrey}{colorModel0)}^;
  ^\color{darkgrey}{field0}^.paintComponent(^\color{darkgrey}{graphics2D0}^);
  assertFalse(^\color{darkgrey}{field0}^.isFocusTraversalPolicySet());
}
/* Field class parent*/
class Container{
  public boolean isFocusTraversalPolicySet() {
    return (^\color{darkgrey}{focusTraversalPolicy}^ != null);
  }
}
\end{lstlisting}
\end{figure}
This category is similar to the previous one, except that the methods used in the asserts belong to a parent class of the class under test. We have separated this category as we believe it may be useful for tool maintainers.

\begin{rmd}
Test generator tools should prioritize the relationship between the act and assert steps in the generation algorithm to enhance the effectiveness of generated tests.
\end{rmd}



\subsubsection{Redundant Code}
We discovered pieces of duplicate, redundant, or unnecessary test code that could be eliminated for simplicity or extracted to prevent redundancy.

\myparagraph{Duplicated Setup} We observed that several tests within the same test suite initialize the same group of objects at the beginning. Although not all tests in the suite have the same initialization, some tests contain exactly the same statements at the start. For instance, \ct{test07}, \ct{test08}, and \ct{test09} (shown in Figure \ref{fig:duplicated_setup}) are generated tests that demonstrate this duplication, as each of them initializes the variables \ct{scriptOrFnScope0} and \ct{scriptOrFnScope1} using the same constructor and arguments. Therefore, the duplicated code could be extracted into a setUp method to avoid redundancy.

\begin{figure}
\begin{lstlisting}[language = Java, escapechar=^, label={fig:duplicated_setup}, caption = Example of test with duplicated setup.]
public void test07() throws Throwable {
  ScriptOrFnScope ^\color{darkgrey}{scriptOrFnScope0}^ = 
    new ScriptOrFnScope((-806), (ScriptOrFnScope) null);
  ScriptOrFnScope ^\color{darkgrey}{scriptOrFnScope1}^ = 
    new ScriptOrFnScope((-330), ^\color{darkgrey}{scriptOrFnScope0}^);
  ^\color{darkgrey}{scriptOrFnScope1}^.preventMunging();
  ...
}
public void test08() throws Throwable {
  ScriptOrFnScope ^\color{darkgrey}{scriptOrFnScope0}^ = 
    new ScriptOrFnScope((-806), (ScriptOrFnScope) null);
  ScriptOrFnScope ^\color{darkgrey}{scriptOrFnScope1}^ = 
    new ScriptOrFnScope((-330), ^\color{darkgrey}{scriptOrFnScope0}^);
  ^\color{darkgrey}{scriptOrFnScope1}^.declareIdentifier("");
  ...
}
public void test09() throws Throwable {
  ScriptOrFnScope ^\color{darkgrey}{scriptOrFnScope0}^ = 
    new ScriptOrFnScope((-806), (ScriptOrFnScope) null);
  ScriptOrFnScope ^\color{darkgrey}{scriptOrFnScope1}^ = 
    new ScriptOrFnScope((-330), ^\color{darkgrey}{scriptOrFnScope0}^);
  ^\color{darkgrey}{scriptOrFnScope1}^.munge();
  ...
}
\end{lstlisting}
\end{figure}
We found 498 (21.2\%) test methods with at least two identical setup statements. These test methods belong to 35 test classes. It is important to note that \emph{EvoSuite} generates a test class for each class under test, which means that there are at least 14 test methods per test class on average.

\begin{rmd}
Test generator tools should identify and refactor tests with duplicated setups into a separate class with a common setup method. The latter will reduce code duplication and improve code quality.
\end{rmd}



\myparagraph{Testing the same exception scenario} We found 166 (7.1\%) tests that test the same exception scenario as another test in the battery, with minor differences. These tests contain mostly the same statements, differing only in the method call that triggers the exception. However, the root cause of the exception is either related to the shared statements or another common factor. For example, two generated tests (\ct{test04} and \ct{test08}) differ only in the method called (\ct{initialize} and \ct{getClusterNodeAddresses}), but both test the same exception scenario where \ct{PropsValues} cannot be found (shown in Figure \ref{fig:testing_same_exception_scenario}). As such, we consider one of these tests redundant. This category of tests handles various exception kinds, such as \ct{NullPointerException}, \ct{RuntimeException}, and \ct{ClassCastException}.


\begin{figure}
 \begin{lstlisting}[language = Java , ,escapechar=^, label={fig:testing_same_exception_scenario}, caption = Testing the same exception scenario.]
public void test04() throws Throwable {
  ClusterExecutorUtil ^\color{darkgrey}{clusterExecutorUtil0}^ = 
    new ClusterExecutorUtil();
  ClusterExecutorImpl ^\color{darkgrey}{clusterExecutorImpl0}^ = 
    new ClusterExecutorImpl();
  ^\color{darkgrey}{clusterExecutorUtil0}^.setClusterExecutor(^\color{darkgrey}{clusterExecutorImpl0}^);
  // Undeclared exception!
  try { 
    ClusterExecutorUtil.initialize();
    fail("Expecting exception: NoClassDefFoundError");
  } catch(NoClassDefFoundError ^\color{darkgrey}{e}^) {
    // Could not initialize class com.liferay.portal.util.PropsValues
    verifyException("com.liferay.portal.cluster.ClusterBase", ^\color{darkgrey}{e}^);
  }
}
public void test08() throws Throwable {
  ClusterExecutorUtil ^\color{darkgrey}{clusterExecutorUtil0}^ = 
    new ClusterExecutorUtil();
  ClusterExecutorImpl ^\color{darkgrey}{clusterExecutorImpl0}^ = 
    new ClusterExecutorImpl();
  ^\color{darkgrey}{clusterExecutorUtil0}^.setClusterExecutor(^\color{darkgrey}{clusterExecutorImpl0}^);
  // Undeclared exception!
  try { 
   ClusterExecutorUtil.getClusterNodeAddresses();
   fail("Expecting exception: NoClassDefFoundError");
  } catch(NoClassDefFoundError ^\color{darkgrey}{e}^) {
   // Could not initialize class com.liferay.portal.util.PropsValues
   verifyException(
     "com.liferay.portal.cluster.ClusterBase", ^\color{darkgrey}{e}^);
  }
}
\end{lstlisting}
\end{figure}


\begin{rmd}
To improve the quality of the test battery and reduce its size, test generator tools should detect and eliminate redundant tests using inter-test analysis, especially for exception scenarios.
\end{rmd}

\myparagraph{Testing the same void method} We found 117 (5\%) tests that evaluate the same void method with small variations. While these variations may help cover new branches of the class under test, the assertions of these tests are unrelated to the branch under test, limiting their usefulness for bug fault detection. For example, consider two generated tests (Figure \coderef{fig:multiple_calls_to_the_same_void_method}). In both tests, the \ct{release} method is called from \ct{DbConnectionBroker}, which affects the \ct{maxConnections} attribute. However, the tests assert a different attribute (\ct{maxReached}) and do not verify the method's side effects. This category is similar to the previous category, "not asserted side-effect," but with two or more tests executing different branches of the same void method, where the assertions are unrelated to this method, further limiting their usefulness. 

\begin{figure}
\begin{lstlisting}[language = Java , escapechar=^, label={fig:multiple_calls_to_the_same_void_method}, caption = Multiple calls to the same void method.]
public void test00() throws Throwable {
  DbConnectionBroker ^\color{darkgrey}{dbConnectionBroker0}^ = new DbConnectionBroker();
  DbConnectionAttributes ^\color{darkgrey}{dbConnectionAttributes0}^ = new DbConnectionAttributes(10);
  ^\color{darkgrey}{dbConnectionBroker0}^.release(^\color{darkgrey}{dbConnectionAttributes0}^);
  assertEquals(0, ^\color{darkgrey}{dbConnectionBroker0}^.getMaxReached());
}
public void test11() throws Throwable {
  DbConnectionBroker ^\color{darkgrey}{dbConnectionBroker0}^ = new DbConnectionBroker();
  assertEquals(30, ^\color{darkgrey}{dbConnectionBroker0}^.getMax());
  DbConnectionAttributes ^\color{darkgrey}{dbConnectionAttributes0}^ = new DbConnectionAttributes(15);
  ^\color{darkgrey}{dbConnectionBroker0}^.release(^\color{darkgrey}{dbConnectionAttributes0}^);
  assertEquals(0, ^\color{darkgrey}{dbConnectionBroker0}^.getMaxReached());
}
\end{lstlisting}
\end{figure}

\begin{rmd}
Test generator tools should generate assertions that verify the behavior associated with different branches of the same method already covered by the test battery.
\end{rmd}

\myparagraph{Asserting object initialization multiple times}
This category includes tests with assertions that verify behavior related to the object initialization during the test setup phase. While testing for proper initialization is valid, some tests in the same battery focus solely on initialization instead of the tested method. These tests often have similar assertions. For example, \ct{test14} and \ct{test27} in Figure \ref{fig:asserting_object_initialization_multiple_times} have identical assertions for the \ct{PhotoController} class that verify values set during initialization.  Note that although these two tests differ slightly, with \ct{test14} calling the method \ct{setLens} and \ct{test27} calling the method \ct{getContentManager}, there are no assertions related to these methods. We found 550 (23,5\%) tests that assert the object initialization multiple times. These tests belong to 50 test suites.



\begin{figure}
\begin{lstlisting}[language = Java, escapechar=^ , label={fig:asserting_object_initialization_multiple_times}, caption = Asserting object initialization multiple times.]
public void test14() throws Throwable {
  Home home0 = new Home();
  SwingViewFactory swingViewFactory0 = new SwingViewFactory();
  PhotoController photoController0 = new PhotoController(home0, (UserPreferences) null, (View) null, swingViewFactory0, (ContentManager) null);
  Camera.Lens camera_Lens0 = Camera.Lens.PINHOLE;
  photoController0.setLens(camera_Lens0);
  assertEquals(0, photoController0.getQuality());
  assertEquals(300, photoController0.getHeight());
  assertEquals(400, photoController0.getWidth());
  assertEquals(1.0F, photoController0.get3DViewAspectRatio(), 0.01F);
  assertEquals(1392409281320L, photoController0.getTime());
  assertEquals(13684944, photoController0.getCeilingLightColor());
}
public void test27() throws Throwable {
  Home home0 = new Home();
  SwingViewFactory swingViewFactory0 = new SwingViewFactory();
  PhotoController photoController0 = new PhotoController(home0,(UserPreferences)null(View) null, swingViewFactory0, (ContentManager) null);
  photoController0.getContentManager();
  assertEquals(13684944, photoController0.getCeilingLightColor());
  assertEquals(300, photoController0.getHeight());
  assertEquals(0, photoController0.getQuality());
  assertEquals(400, photoController0.getWidth());
  assertEquals(1392409281320L, photoController0.getTime());
  assertEquals(1.0F, photoController0.get3DViewAspectRatio(), 0.01F);
}
\end{lstlisting}
\end{figure}

\begin{rmd}
Test generator tools should avoid redundant tests that verify the same object initialization in multiple tests and instead focus on testing the particular behavior of each test. This approach can reduce the size of the test suite and ensure that each test contributes uniquely to code coverage and error detection.
\end{rmd}

\myparagraph{Redundant not Null Assertion}
We found 77 (3.3\%) tests with redundant not null assertions. In particular, we identify two cases: \emph{(i) When the not null assertion is used to verify a recently created object --} For example, consider the test generated for the class \ct{HookHotDeployListener} in Figure \ref{fig:assert-not-null-object-recently-initialized}. The test initializes an instance of the class using the \ct{new} operator and immediately follows it with an \ct{assertNotNull} statement over the newly created object. However, since an error would be triggered if the object were not created, verifying its existence with an \ct{assertNotNull} statement is unnecessary.

\begin{figure}
\begin{lstlisting}[language = Java, escapechar=^ , label={fig:assert-not-null-object-recently-initialized}, caption = Redundant not null assertion that verify a recently created object]
public void test00() throws Throwable {
  HookHotDeployListener ^\color{darkgrey}{hookHotDeployListener0}^ = new HookHotDeployListener();
  assertNotNull(^\color{darkgrey}{hookHotDeployListener0}^);    
}
\end{lstlisting}
\end{figure}



 % ...

\emph{(ii) When another assertion indirectly verifies that a given variable is not null, and deleting will not alter the test semantics. --} This happens when a test method contains another assertion that indirectly verifies that a given variable is not null. Therefore the \ct{assertNotNull} could be deleted without changing the test behavior\footnote{It is important to mention that there are cases where even though another assertion indirectly verifies a variable is not null, the not null assertion is not redundant.}. To illustrate, consider the following generated test for the class \ct{SubstringLabeler}, where there is an \ct{assertNotNull} for the variable \ct{string0} and an \ct{assertEquals} that checks that the variable \ct{string0} is equal to the string \ct{"trainingSet"} (Figure \ref{fig:assert-not-null-object-redundant-assert}). In this test, the \ct{assertEquals} indirectly verifies that the variable \ct{string0} is not null since it does not contain the expected value. 

\begin{figure}
\begin{lstlisting}[language = Java, escapechar=^ , label={fig:assert-not-null-object-redundant-assert}, caption = Redundant not null assertion since the assertEquals indirectly \\verifies that the variable is not null]
public void test74() throws Throwable {
  ...
  String string0 = ^\color{darkgrey}{substringLabeler\_Match0}^.getMatch();
  assertNotNull(^\color{darkgrey}{string0}^);
  assertEquals("trainingSet", ^\color{darkgrey}{string0}^);
  ...
}
\end{lstlisting}
\end{figure}
We found 47 tests that contained an \ct{assertNotNull} after initializing an object and 33 tests that presented another assertion that indirectly verify the content of a given variable besides the \ct{assertNotNull}.

\begin{rmd}
To improve the effectiveness of test generation, it is recommended that test generators analyze the need for not-null assertions and avoid generating redundant assertions.
\end{rmd}


\subsubsection{Failed Setup}
Previous studies show that many tests require significant setup but often fail, causing exceptions. If the test generator tool struggles with initializing objects, it can result in a large number of tests with failed setups. We identified three types of failed setups during our manual inspection, including eight test suites with all test methods handling exceptions due to failed setups.
\myparagraph{Exceptions due to null arguments}
We observed that 300 (12.8\%) tests trigger a \ct{NullPointerException} because null arguments are sent to a method call or class constructor. For example, consider the \ct{test02} in Figure \ref{fig:exceptions-due-to-null-arguments}, where a null value is sent to the \ct{refresh} method. The exception is directly linked to the null value and is triggered when the \ct{refresh} method attempts to execute the \ct{getDataSource} method on the null value. Note that the \ct{refresh} method does not handle null arguments. While sending a null argument may help cover new branches or test exception scenarios, these tests have limited usefulness as the exception triggered is directly linked to the null value rather than any new branch of the tested method.

\begin{figure}
 \begin{lstlisting}[language = Java, escapechar=^ , label={fig:exceptions-due-to-null-arguments}, caption = Exceptions due to null arguments.]
public void test02() throws Throwable {
  PeersItem ^\color{darkgrey}{peersItem0}^ = new PeersItem();
  // Undeclared exception!
  try { 
    ^\color{darkgrey}{peersItem0}^.refresh((TableCell) null);
    fail("Expecting exception: NullPointerException");
  } catch(NullPointerException ^\color{darkgrey}{e}^) {
  // no message in exception (getMessage() returned null)      
    verifyException("org.gudy.azureus2.ui.swt.views.tableitems.tracker.PeersItem", ^\color{darkgrey}{e}^);
  }
}
\end{lstlisting}
\end{figure}

\myparagraph{Exceptions due external dependencies} 
This category involves tests that handle exceptions triggered by external dependencies or resources rather than due to issues in the behavior of the method being called. For example, consider the following generated test for the \ct{EntryListView} class shown in Figure \ref{fig:exceptions-due-to-external-dependencies}, which includes a try-catch statement that catches a \ct{HeadlessException}. This exception type is linked to code that relies on keyboard, display, or mouse events in an environment that does not support them. We found similar exceptions such as \ct{SQLException} and \ct{NotYetConnectedException}. We found 52 (2.2\%) test methods that handle exceptions due to external dependencies.
\begin{figure}
 \begin{lstlisting}[language = Java, escapechar=^ , label={fig:exceptions-due-to-external-dependencies}, caption = Exceptions due to external dependencies.]
public void test00() throws Throwable {
  try { 
    EntryListView.showView();
    fail("Expecting exception: HeadlessException");
  } catch(HeadlessException ^\color{darkgrey}{e}^) {
// no message in exception (getMessage() returned null)
     verifyException("java.awt.GraphicsEnvironment", ^\color{darkgrey}{e}^);
  }
}
\end{lstlisting}
\end{figure}
Multiple tests in the same test class have similar exception handling code and the same root cause, which is a failed setup. 



\myparagraph{Exceptions due to incomplete setup}
This category contains tests that handle exceptions triggered by an incomplete setup. We analyzed the identified cases and discovered that uninitialized values or uncalled methods triggered the exceptions. To illustrate, consider the following generated test for the \ct{AlphabeticTokenizer} class shown in Figure \ref{fig:exceptions-due-to-incomplete-call-sequence}. To successfully call the \ct{nextElement} method, the \ct{m_CurrentPos} attribute needs to be initialized. Otherwise, the method will attempt to access a null index inside an array. In \ct{test00}, the default constructor of \ct{AlphabeticTokenizer0} does not initialize \ct{m_CurrentPos}. Therefore, when calling \ct{nextElement}, a \ct{NullPointerException} is thrown. We found 41 (1.7\%) test methods that handle exceptions due to external dependencies.

\begin{figure}
 \begin{lstlisting}[language = Java , escapechar=^,  label={fig:exceptions-due-to-incomplete-call-sequence}, caption = Exceptions due to incomplete call sequence.]
public void test00() throws Throwable {
  AlphabeticTokenizer ^\color{darkgrey}{alphabeticTokenizer0}^ = new AlphabeticTokenizer();
  // Undeclared exception!
  try { 
    ^\color{darkgrey}{alphabeticTokenizer0}^.nextElement();
    fail("Expecting exception: NullPointerException");
  } catch(NullPointerException ^\color{darkgrey}{e}^) {
    // no message in exception (getMessage() returned null)
    verifyException("weka.core.tokenizers.AlphabeticTokenizer", ^\color{darkgrey}{e}^);
  }
}
/* Class under test*/
class AlphabeticTokenizer{
  protected int m_CurrentPos;
  public Object nextElement() {
    int beginpos, endpos;       
    beginpos = m_CurrentPos;
    while ( (beginpos < m_Str.length) && ((m_Str[beginpos] < 'a') && (m_Str[beginpos] > 'z')) && ((m_Str[beginpos] < 'A') && (m_Str[beginpos] > 'Z')) ) {
      beginpos++;    
    }
    ...
    return s;
  }
}
\end{lstlisting}
\end{figure}

\begin{rmd}
Test generator tools should be able to detect situations where the object initialization is unsuccessful because this can lead to multiple tests with similar exception scenarios.
\end{rmd}

\subsubsection{Testing Field Accessors, and Constants}
Tests, where the behavior under test is rather simple, belong to this category. Apart from improving code coverage and preventing exceptions, the behavior under test is predictable and straightforward.

\myparagraph{Testing only field accessors} 
We identified 139 (5.9\%) tests in which, apart from object initialization, the behavior under test solely consisted of field accessors such as getters and setters. An example of this is the generated test for the \ct{DirEntry} class shown in \coderef{fig:testing_only_field_accessors}, which solely reads and writes the \ct{modtime} field using the \ct{getModtime} and \ct{setModtime} methods. Although calling these methods increases coverage for the class under test, we believe that testing only field accessors that do not add any additional behavior beyond exposing a class field are strong candidates for test reduction or prioritization.
 \begin{figure}
\begin{lstlisting}[language = Java, escapechar=^ ,label={fig:testing_only_field_accessors}, caption = Testing only field accessors.]
public void test05() throws Throwable {
  DirEntry ^\color{darkgrey}{dirEntry0}^ = new DirEntry();
  ^\color{darkgrey}{dirEntry0}^.setModtime("");
  String ^\color{darkgrey}{string0}^ = ^\color{darkgrey}{dirEntry0}^.getModtime();
  assertEquals("", ^\color{darkgrey}{string0}^);
}
\end{lstlisting}
\end{figure}

\myparagraph{Asserting Constants} 
We observed that 20 (0.8\%) tests only check the value of a constant using assertions. While testing constant values can be valuable in certain cases, these tests also tend to call different methods and only assert unrelated constant values. For example, consider the generated test for the \ct{ConnectionConsumer} class shown in \coderef{fig:asserting_constants}. This test initializes an object of the class and invokes the \ct{connectionError} method, but the only assertion in the test checks the \ct{DEFAULT_SUBSEQUENT_RETRIES} constant, which is always initialized to zero unless manually modified. Moreover, this constant is irrelevant to the \ct{connectionError} method. Tests that only check constant values in this manner do not significantly contribute to the test suite, as the outcome of such assertions is always obvious. 
\begin{figure}
\begin{lstlisting}[language = Java, escapechar=^ ,label={fig:asserting_constants} , caption = Asserting constants.]
public void test06() throws Throwable {
  ...
  ^\color{darkgrey}{connectionConsumer0}^.connectionError("summa.persistent");
  assertEquals(0, ConnectionConsumer.DEFAULT_SUBSEQUENT_RETRIES);
}
\end{lstlisting}
\end{figure}
\begin{rmd}
Tests that solely use accessors or validate constant values might not substantially impact the test suite. Therefore, test generator tools should identify such tests as potential candidates for prioritization or removal to reduce the total number of tests to execute or analyze.
\end{rmd}


\subsection{Quality Issue Distribution (RQ2)}
The frequency of the categories above at the test method, class, and project levels can be found in Table \ref{tab:category-frequency}. Our results indicate that the reported issues were present across various classes and projects within the data set. The latter suggests that the issues are not specific to any particular project or class. Although the percentage of test methods with at least one issue is small (less than 20\%), the number of classes and test suites containing at least one issue is considerable. For instance, the issue ``not asserted side effects'' appear at least once in 57 out of 100 generated test suites.

\begin{table}[!ht]
\scriptsize
\centering
\label{tab:category-frequency}
\caption{Fine-grained Quality Issues: Categories and Frequency}
\setlength{\tabcolsep}{1.8pt}
\renewcommand{\arraystretch}{1.2}
\begin{tabular}{rp{4.5cm}rrrr}
\multicolumn{2}{l}{\textbf{}}                                                         
& \multicolumn{1}{r}{\setlength\extrarowheight{-3pt}\textbf{\begin{tabular}[c]{@{}c@{}}\% Test\\ Methods\end{tabular}}}
& \multicolumn{1}{r}{\setlength\extrarowheight{-3pt}\textbf{\begin{tabular}[c]{@{}c@{}}\# Test\\ Methods\end{tabular}}}
& \multicolumn{1}{r}{\setlength\extrarowheight{-3pt}\textbf{\begin{tabular}[c]{@{}c@{}}\# Test\\ Classes\end{tabular}}} 
& \multicolumn{1}{r}{\setlength\extrarowheight{-3pt}\textbf{\begin{tabular}[c]{@{}c@{}}\# \\ Projects\end{tabular}}} 

\\ \hline 
\multicolumn{6}{|l|}{\cellcolor[HTML]{EFEFEF}\textbf{Act-Assert Mismatch}} \\ \hline
  & Not asserted side effects                                        & 14.0\% & 328    & 49 &  26 \\ \cline{2-6}
  & Not asserted return values                                       & 10,3\%  & 242     & 30 &  19 \\ \cline{2-6}
  & Assertions w/ not related parent class method                  & 5,1\%  & 121     & 13 & 9 \\ \hline 
\multicolumn{6}{|l|}{\cellcolor[HTML]{EFEFEF}\textbf{Redundant Code}} \\  \hline  
 & Asserting object initialization multiple times                   & 23,5\% & 550   & 50  &  24 \\ \cline{2-6}
 & Duplicated setup                                                 & 21,2\% & 498   & 35  & 18 \\ \cline{2-6}
 & Testing the same exception  scenario                             & 7,1\%  & 166   & 24  & 16  \\ \cline{2-6}
 & Testing the same void method                         & 5,0\%  & 117   & 20  & 15 \\ \cline{2-6}
 & Redundant not null assertion                                               & 3,3\%  & 77   & 12  & 12 \\ \hline 
\multicolumn{6}{|l|}{\cellcolor[HTML]{EFEFEF}\textbf{Failed Setup}}        \\ \hline 
 
 & Exceptions due to null arguments                          & 12,8\% & 300         & 64 & 30 \\ \cline{2-6}
 & Exceptions due to incomplete setup                               & 2,2\%  & 52         & 19 & 12 \\ \cline{2-6}
 & Exceptions due to external dependencies                          & 1,7\%  & 41         & 20 & 12 \\ \hline
 \multicolumn{6}{|l|}{\cellcolor[HTML]{EFEFEF}\textbf{Testing Field Accessors and Constants}} \\ \hline 
 & Testing only field accesors                                      & 5,9\%   & 139     & 17 & 13 \\ \cline{2-6}
 & Asserting constants                                              & 0,8\%   & 20      & 11 & 6 \\ \hline 
\end{tabular}
\end{table}

\begin{figure}[!ht]
\centering
\setcounter{figure}{15}
\includegraphics[scale=0.32]{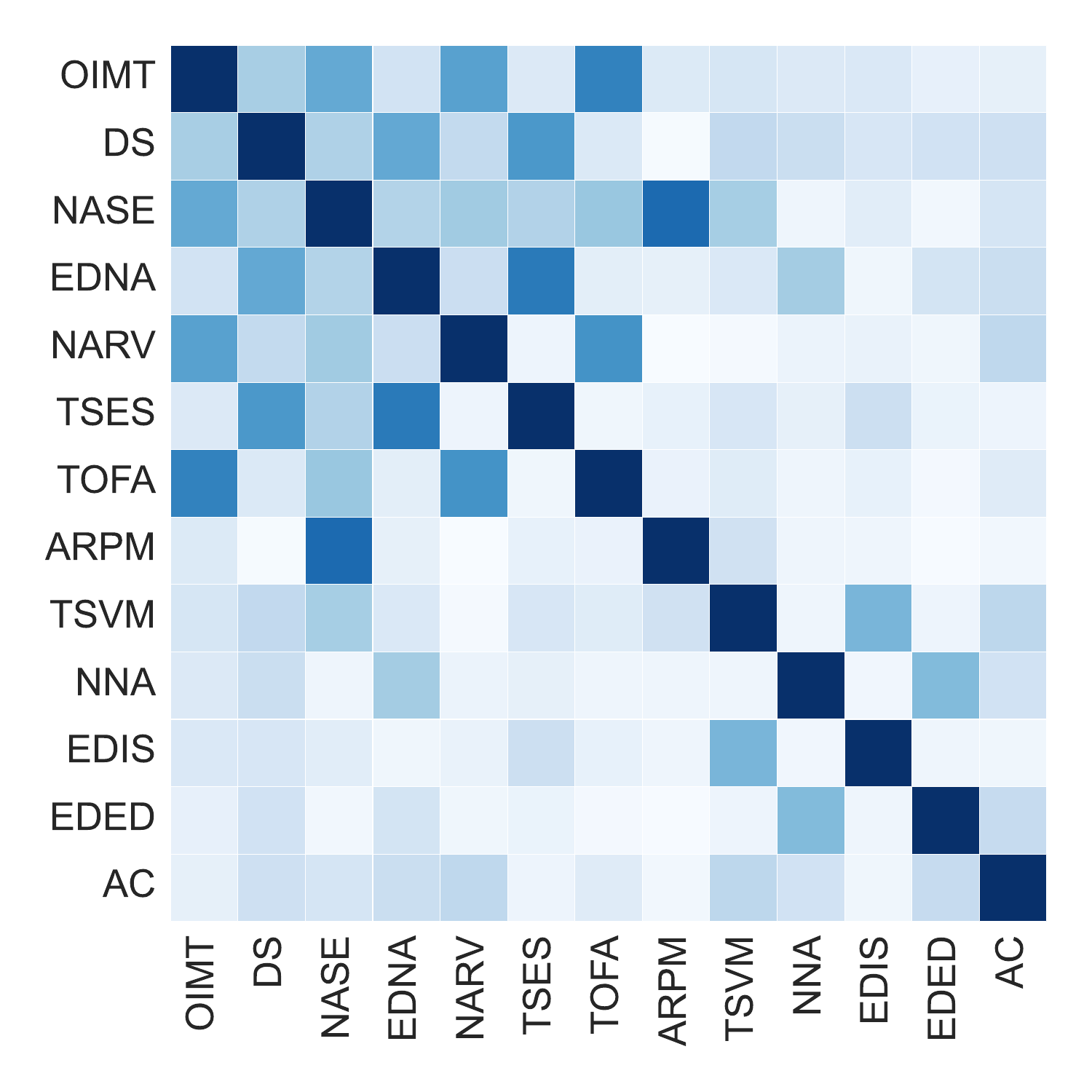}
\caption{Relation between identified quality issues}
\label{fig:heatmapSquare}
\end{figure}

\myparagraph{Relation between Quality Issues} The heat map shown in \figref{heatmapSquare} shows the relationships between the identified issues, with the color of each cell indicating the number of generated tests that have both of these issues. The two most closely related quality issues are \emph{"Not asserted side effects" (NASE)} and \emph{"Assertion with no related parent class" (ARPM)}. Therefore, a significant proportion of tests that do not assert side effects also contain an assertion with a method call to a parent class unrelated to the under-test behavior. Two other related issues are: \emph{"Exceptions due to null arguments" (TWNA)} and \emph{"Testing the same exception scenario" (TSES)}. Although the tests with these issues have slightly different codes, in many cases, the root cause of the exception is null arguments. It is worth noting that the exception scenarios are not exclusively related to null values, which is why the color box is not dark blue. Overall, \figref{heatmapSquare} shows that no two quality issues overlap entirely in the same group of tests, suggesting that each issue has unique characteristics not covered by other issues.

\section{Threats to Validity}
This study presents a manual categorization of a particular sample of automatically generated tests, and like any other study, it is subject to a number of threats to validity.

\myparagraph{Constructor Validity} In order to search for code parts or code anti-patterns that do not match the current test smell definitions, we use a set of test smells as a basis for the analysis. The fact that not all relevant test smells proposed by previous literature are covered may be a threat to the validity of this study. Consequently, a quality issue may partly overlap the definition of a test smell not included in our set of test smells. To reduce this threat, we utilized the test smells considered by Panichella and colleagues~\cite{panichella2022}, since this study collects the test smells proposed by formally-published sources~\cite{peruma2019, Grano2018,deursen2001} which can affect  tests generated by EvoSuite.

\myparagraph{Internal Validity} We performed a thematic analysis to answer our research questions. Two authors performed the systematic process to conduct the analysis. Since this process includes generating codes and defining themes, these aspects vary depending on the coder’s experience, point of view, and level of abstraction. For example, to respond to RQ1, we detected quality issues involved with duplicated, redundant, and/or unnecessary test code. These quality issues are related to the code smell of duplicated code. However, each quality issue shows particularities associated to the nature of generated tests. We tried to reduce this threat by checking the consistency of the process. The remaining authors examined the description of themes, generated codes, and the test cases involved with codes. We held three discussion meetings to diagnose the codes and the themes generated. As a result, we solve any discrepancies.

\myparagraph{External Validity} In our study, we identified and categorized quality issues in a sample of 2,340 tests generated by EvoSuite. However, besides involving a significant manual effort, the generalizability of our findings is limited. Additional issues may be present in tests generated by other tools or in different projects. Nevertheless, this paper presents an initial taxonomy of quality issues highlighting the discrepancy between generated and manually-written tests. The latter confirms that test smells alone do not capture all the quality issues present in generated tests.

\section{Conclusion}

This paper presents a manual categorization of quality issues for a data set of 2,340 generated tests, resulting in a taxonomy of 13 issues that reflect atypical patterns in automatically generated tests. Our study shows that the reported issues are not specific to a particular class or project. Furthermore, our issue catalog highlights the importance of considering the code being tested when analyzing the test quality. For example, methods with side effects and complex object initialization  may trigger a number of issues in the generated tests. Our findings are consistent with previous work in revealing a discrepancy between the optimization metric used by test generator tools and the usefulness of the generated tests. Based on the issues we identified, we provide eight recommendations that demonstrate their applicability and usefulness. As future work, we plan to extend the external validity of the issues found by automatically searching the reported issues in other projects, and generator tools.

\section*{Acknowledgment}

Juan Pablo Sandoval Alcocer thanks ANID FONDECYT Iniciación Folio 11220885 for supporting this article.

\bibliographystyle{plain}
\bibliography{references}

\end{document}